\documentclass[reprint,superscriptaddress, amsmath,amssymb, aps]{revtex4-2}

\usepackage{color}
\usepackage{graphicx}
\usepackage{dcolumn}
\usepackage{bm}
\usepackage{hyperref}
\usepackage{footmisc}
\usepackage{comment} 
\usepackage{braket}
\usepackage{mathtools}
\usepackage{orcidlink}

\newcommand{\calH}{\mathcal{H}}

\newcommand{\eff}{\mathrm{eff}}

\begin{document}

\title{Ferromagnetism in tilted fermionic Mott insulators}

\author{Kazuaki Takasan \orcidlink{0000-0002-7827-1690}}
\email{kazuaki.takasan@phys.s.u-tokyo.ac.jp}
\affiliation{
Department of Physics, University of California, Berkeley, California 94720, USA}
\affiliation{
Materials Sciences Division, Lawrence Berkeley National Laboratory, Berkeley, California 94720, USA
}
\affiliation{
Department of Physics, University of Tokyo, 7-3-1 Hongo, Tokyo 113-0033, Japan}

\author{Masaki Tezuka \orcidlink{0000-0001-7877-0839}}
\email{tezuka@scphys.kyoto-u.ac.jp}
\affiliation{
Department of Physics, Kyoto University, Kyoto 606-8502, Japan
}
\date{\today}
             
\begin{abstract}
We investigate the magnetism in tilted fermionic Mott insulators. With a small tilt, the fermions are still localized and form a Mott-insulating state, where the localized spins interact via antiferromagnetic exchange coupling. While the localized state is naively expected to be broken with a large tilt, in fact, the fermions are still localized under a large tilt due to the Wannier-Stark localization and it can be regarded as a localized spin system. We find that the sign of the exchange coupling is changed and the ferromagnetic interaction is realized under the large tilt. To show this, we employ perturbation theory and real-time numerical simulation of the fermionic Hubbard chain. Our simulation exhibits that it is possible to effectively control the speed and time direction of the dynamics by changing the size of the tilt, which may be useful for experimentally measuring out-of-time-ordered correlators. Finally, we discuss experimental platforms, such as ultracold atoms in an optical lattice, to observe these phenomena.
\end{abstract}

\maketitle

\section{Introduction}
The effect of a tilted potential in periodic systems has been studied for a long time. This is because a linear potential corresponds to a static electric field, and electric-field effects in solids are an important issue in condensed matter physics from both fundamental and applied viewpoints. In this direction, various interesting and significant phenomena, such as Bloch oscillations~\cite{Bloch1929}, Zener tunneling~\cite{Zener1934}, and Wannier-Stark ladders~\cite{Wannier1960, Gluck2002}, have been established. While the understanding of these phenomena has been largely advanced, their counterparts in strongly interacting systems are still not fully understood and remain an intriguing topic~\cite{Taguchi2000, Yamakawa2017, Oka2003, Eckstein2010, Meisner2010, Oka2012, Aron2012, Takasan2019breakdown}. A prominent example is the breakdown of Mott insulators under strong electric fields (dielectric breakdown), which has been studied theoretically and experimentally~\cite{Taguchi2000, Yamakawa2017, Oka2003, Eckstein2010, Meisner2010, Oka2012, Aron2012}. A recent review on nonequilibrium phenomena in electric-field-driven Mott insulators is given in Ref.~\cite{Murakami2025}.

A tilted potential is also realized in atomic-molecular-optical (AMO) systems such as ultracold atoms and is used as a convenient tool to induce or control various quantum many-body phenomena. For instance, bosonic Bloch oscillations~\cite{Kolovsky2004, Mahmud2014, Meinert2014, Geiger2018} and Zener tunneling~\cite{Tomadin2008, Chen2011, Kolovsky2016LZ} have been explored. Quantum phase transitions induced by a tilt have also been widely investigated~\cite{Sachdev2002, Pielawa2011, Simon2011, Kolodrubetz2012, Kolovsky2016, Buyskikh2019}. In recent years, tilted potentials have attracted growing attention in the context of thermalization in closed quantum systems~\cite{Schulz2019, vanNieuwenburg2019, Sala2020, Khemani2020, Doggen2021, Desaules2021, GuardadoSanchez2020, Scherg2021, Kohlert2021, Morong2021, Guo2020, Adler2024}. Interacting fermions in a tilted lattice have been found to show behavior similar to many-body localization in disordered systems~\cite{Schulz2019, vanNieuwenburg2019}, known as Stark many-body localization. This phenomenon and related effects, including Hilbert-space fragmentation and shattering~\cite{Sala2020, Khemani2020, Doggen2021}, as well as quantum many-body scars in tilted Hubbard models~\cite{Desaules2021}, have been studied extensively. Furthermore, related experiments in ultracold atoms~\cite{GuardadoSanchez2020, Scherg2021, Kohlert2021, Adler2024}, trapped ions~\cite{Morong2021}, and superconducting qubits~\cite{Guo2020} have demonstrated that tilted-potential systems constitute an important platform for investigating quantum many-body phenomena.

\begin{figure}
    \includegraphics[width=8.5cm]{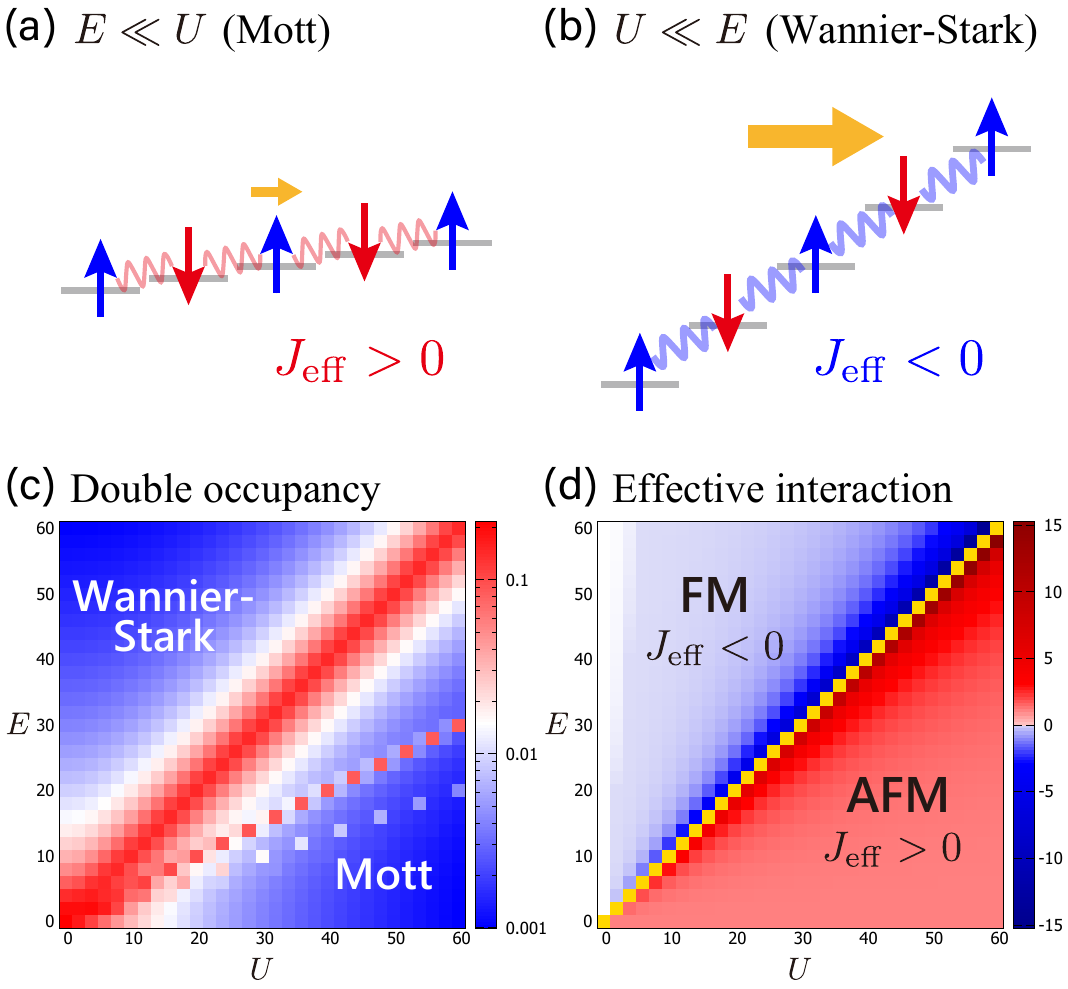}
    \caption{
    (a, b) Schematic picture of (a) the Mott insulating regime ($E \ll U$) and (b) the Wannier-Stark localized regime ($U \ll E$). (c) Time average of the double occupancy per site $\overline{\mathcal{N}}_\mathrm{double}$ [Eq.~\eqref{eq:ave_double_occ}], starting from the singly occupied state $\ket{\uparrow\downarrow\uparrow\downarrow\cdots}$ for $L=10$. The time step $\Delta t$ is set to $1/800$. $\overline{\mathcal{N}}_\mathrm{double}$ takes a large value around the resonant condition $nE=U$ ($n=1,2,\cdots$) where the insulating state easily breaks down.
    (d) Effective interaction $J_\mathrm{eff}(E)/J_\mathrm{eff}(0)$ defined in Eq.~\eqref{eq:J_eff}. $t_h$ ($t_h^{-1}$) is used as the unit of energy (time).
    }
    \label{fig:fig1}
\end{figure}

In this paper, we investigate magnetism, which is one of the most important quantum many-body phenomena, in tilted fermionic Mott insulators. This is experimentally relevant to both Mott insulator materials under a static electric field and AMO systems with a linear potential. Despite its importance and experimental relevance, most previous studies have focused on charge dynamics, localization phenomena, or constrained Hilbert-space structures, and the role of magnetic interactions in tilted fermionic Mott insulators has remained relatively unexplored. One of the authors of this paper has studied the magnetism of fermionic Mott insulators with a small tilt, schematically shown in Fig.~\ref{fig:fig1}(a)~\cite{Takasan2019}. In Ref.~\cite{Takasan2019}, it was demonstrated that the antiferromagnetic coupling is enhanced with a tilt, which was used for controlling various magnetic phases with electric fields. Recently, this idea has been shown to be applicable to more generic setups and useful for controlling other types of magnetic interactions~\cite{Furuya2021a, Furuya2021b}. In this paper, we address a broader  parameter range of tilt including a much larger one. With a large tilt, it is naively expected that the Mott-insulating state is broken through the many-body Zener breakdown~\cite{Taguchi2000, Yamakawa2017, Oka2003, Eckstein2010, Meisner2010, Oka2012, Aron2012, Takasan2019breakdown}. This is true for the size of the tilt per site at the same order as the on-site interaction. However, with a much larger tilt, the fermions can be localized even under the tilt. This is induced by the Wannier-Stark localization~\cite{Wannier1960, Gluck2002, Schulz2019, vanNieuwenburg2019}, which freezes the charge degree of freedom. Thus, the system is still described as localized spins under a large tilt. Our question is what kind of magnetism emerges in this localized spin system and how the large-tilt regime is connected to the small tilt regime. 

To tackle this issue, we study the one-dimensional Hubbard model with a tilt. One approach is perturbation theory. We derive the effective spin model for the generic size of tilt and find that the ferromagnetic interaction appears in the large tilt regime. The other approach is to solve the many-body Schr\"odinger equation numerically to track the spin dynamics. The numerical result is consistent with perturbation theory. The dynamics under a tilt itself is also interesting because we can control the speed and time direction by changing the size of the tilt. We mention the application of this dynamics to the experimental measurement of out-of-time ordered correlators~\cite{Larkin1969, Maldacena2016}. Finally, we discuss platforms for experimentally observing the signature of ferromagnetism and address several future perspectives.

\section{Model}
We study the one-dimensional fermionic Hubbard model with a linear potential. The Hamiltonian is given by
\begin{align}
    H &= - t_h \sum_{j=1}^{L-1} \sum_{\sigma=\uparrow, \downarrow} ( c^\dagger_{j+1 \sigma} c_{j \sigma} + \mathrm{h.c.}) \nonumber \\
    & \qquad + U \sum_{j=1}^L n_{j \uparrow} n_{j \downarrow}  +\sum_{j=1}^L \sum_{\sigma=\uparrow,\downarrow} jE n_{j \sigma}, \label{eq:model_lgauge}
\end{align}
where $c_{j \sigma}$ ($c_{j \sigma}^\dagger$) is the annihilation (creation) operator of a fermion at the $j$-th site with spin $\sigma (=\uparrow, \downarrow)$ and $n_{j \sigma}=c_{j \sigma}^\dagger c_{j \sigma}$. Here, we choose the open boundary condition, which corresponds to the realistic setup in the AMO systems such as ultracold atoms in an optical lattice~\cite{GuardadoSanchez2020, Scherg2021, Kohlert2021}. $t_h$ and $U$ represent the hopping amplitude and the on-site interaction energy respectively. Throughout this paper, we use $t_h$ ($t_h^{-1}$) as the unit of energy (time). The size of the tilt is denoted by $E$, which is related to the physical electric field $\mathcal{E}$ in electronic systems as $E=|e|a\mathcal{E}/\hbar$ where $e$ and $a$ are the elementary charge and the lattice constant. To study the properties as localized spin systems, we focus on the half-filled and repulsive ($U>0$) case throughout this paper. For the later convenience, we introduce the other gauge choice. With a time-dependent gauge transformation $U(t)=\exp[-iEt\sum_{j, \sigma} j n_{j, \sigma}]$, the Hamiltonian is transformed into $\tilde{H}(t)=U^\dagger H U - i U^\dagger \partial_t U$, which is calculated as
\begin{align}
    \tilde{H}(t) \!
    &= \! - t_h \sum_{j=1}^{L-1} \sum_{\sigma=\uparrow, \downarrow}( e^{iEt} c^\dagger_{j+1 \sigma} c_{j \sigma} + \mathrm{h.c.}) + U \sum_{j=1}^{L} n_{j \uparrow} n_{j \downarrow}. \label{eq:model_vgauge}
\end{align}
This gauge choice is called velocity gauge, whereas the one in Eq.~\eqref{eq:model_lgauge} is called length gauge~\footnote{These two gauges are equivalent in the open boundary condition as far as we study the gauge-invariant quantities. A subtle point arises in the periodic boundary condition. The models \eqref{eq:model_lgauge} and \eqref{eq:model_vgauge} with a periodic boundary condition are not connected with the gauge transformation because of the boundary term. However, as discussed in Appendix \ref{subsec:gaugechoice}, our results do not depend much on the gauge choice even with the periodic boundary conditions.}. 

\begin{figure}
\includegraphics[width=8.5cm]{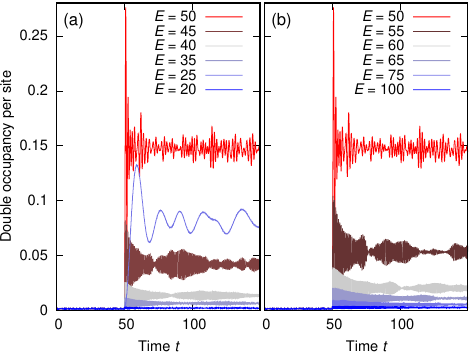}
\caption{Time evolution of the doublon number per site $\mathcal{N}_\mathrm{double}(t)$ [Eq.~\eqref{eq:double_occ}], where electric field $E$ is switched on at $t=50$. The data for $E\leq (\geq)~50$ are shown in the left (right) panel. We set $U=50$, $L=10$, and $\Delta t=1/3200$, and choose the singly occupied state $\vert\uparrow\downarrow\uparrow\downarrow\cdots\uparrow\downarrow\rangle$ as the initial state. $t_h$ ($t_h^{-1}$) is used as the unit of energy (time).
}
\label{fig:doubleocc}
\end{figure}

\section{Localization of the charge degree of freedom}
\label{sec:charge_dynamics}
Let us start by examining the charge dynamics. We numerically solve the many-body Schr\"odinger equation of the model \eqref{eq:model_vgauge} directly using the fourth-order Runge--Kutta method, which can treat only small sizes but provides information independent of approximations as long as we adequately choose the time step $\Delta t$ so that the effect of time discretization is negligible~\cite{FN1}. Note that we use the velocity-gauge Hamiltonian \eqref{eq:model_vgauge} instead of the length-gauge Hamiltonian \eqref{eq:model_lgauge} because of its numerical efficiency~\footnote{The details of this point are addressed in Appendix~\ref{subsec:gaugechoice}.}.  We choose the Ne\'el state $\ket{\uparrow \downarrow \uparrow \downarrow \cdots}$ as the initial state and observe the time evolution of the doublon number per site,  
\begin{gather}
\mathcal{N}_\mathrm{double}(t)=\frac{1}{L}\sum_{j=1}^L \langle\Psi(t)\vert n_{j\uparrow}n_{j\downarrow} \vert\Psi(t)\rangle, \label{eq:double_occ}
\end{gather}
where $\vert\Psi(t)\rangle$ is the many-body wavefunction at time $t$. The field-strength dependence of the doublon dynamics is shown in Fig.~\ref{fig:doubleocc}. For small fields of $E < U$, the doublon number is small under a tilt because the Mott insulating state is still preserved. At the resonant points $nE = U$ ($n=1, 2, \cdots$), the number becomes very large just after applying the electric field, which breaks the Mott insulator. In contrast, for larger values of $E>U$, the double occupancy takes smaller values, of the same order of magnitude as in the Mott insulating regime ($E<U$). This supports the realization of a localized spin system. To obtain the whole picture, we calculate the time average of the doublon number 
\begin{gather}
\overline{\mathcal{N}}_\mathrm{double}=\frac{1}{t_1-t_0} \int_{t_0}^{t_1}dt~\mathcal{N}_\mathrm{double}(t), \label{eq:ave_double_occ}
\end{gather}
with $t_0=10$ and $t_1=100$ when we turn on the tilt at $t=0$~\footnote{Note that this protocol is different from the one in Fig.~\ref{fig:doubleocc} where the tilt is turned on at $t=50$.}. The averaged doublon numbers for different points of $(E, U)$ are summarized in Fig.~\ref{fig:fig1}~(c). This figure shows that we have a well-defined Wannier-Stark localized regime and it has a broad range of parameters. Below, we study the magnetism in the Mott-insulating regime and the Wannier-Stark localized regime where the fermions behave as localized spins.

\section{Effective spin Hamiltonian}
\label{sec:effective_model}
\begin{figure}
    \includegraphics[width=7cm]{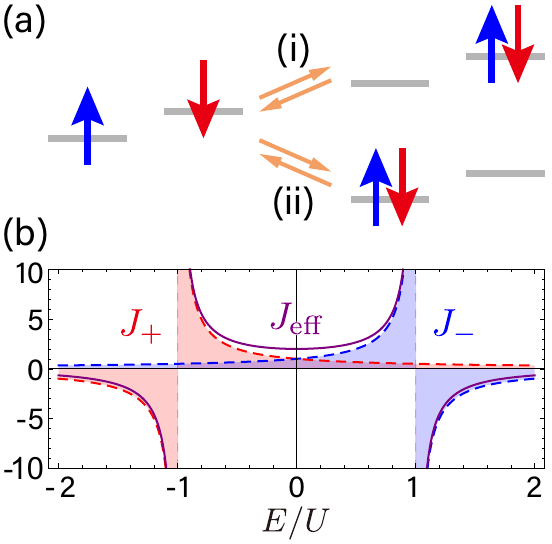}
    \caption{(a) Perturbation processes relevant to the kinetic exchange. The two processes (i) and (ii) are inequivalent only when an electric field is applied. (b) $E$-dependence of $J_\mathrm{\eff}(E)$ [Eq.~\eqref{eq:J_eff}], $J_+(E)$, and $J_-(E)$ [Eq.~\eqref{eq:Jpm}]. The plotted value is normalized by $J_0/2=2t_h^2/U$.}
    \label{fig:perturbation}
\end{figure}

To study the magnetism in the Mott and Wannier-Stark regimes, we start with the effective spin model based on perturbation theory. We consider the strong coupling regime $U \gg t_h$ and treat the hopping term [the first term in Eq.~\eqref{eq:model_lgauge}] as the perturbation. 
Let us begin with the small tilt case $|E|<U$. Here, we assume that the value of $E$ is away from the resonant condition $U=n|E|$ ($n=1, 2, 3, \cdots$), where the double occupancy becomes very large as shown in Fig.~\ref{fig:fig1}~(c) and thus the localized states are broken. In contrast, except for the resonant points, the Mott-insulating state survives even under a small tilt. Thus, the system can be described as localized spins [Fig.~\ref{fig:fig1}~(a)]. The effective model for the spins is the Heisenberg chain with a field-dependent coupling,
\begin{gather}
H_\mathrm{eff}=\sum_{j=1}^L J_\mathrm{eff}(E)\bm S_j \cdot \bm S_{j+1}, \label{eq:H_eff_static} \\
J_\mathrm{eff}(E)=\frac{J_0}{1-\left(\frac{E}{U}\right)^2}, \label{eq:J_eff}
\end{gather}
with $J_0=4t_h^2/U$~\cite{Takasan2019}. The derivation is presented in Appendix \ref{sec:static_derivation}. Here, $\bm S_j = (S^x_j, S^y_j, S^z_j)$ denotes a spin operator at the $j$-th site. Eq.~\eqref{eq:J_eff} shows that the antiferromagnetic exchange coupling is enhanced by adding the tilt~\cite{Takasan2019}. To clarify the physical meaning of Eq.~\eqref{eq:J_eff}, it is useful to decompose it into the following form,
\begin{gather}
J_\mathrm{eff}(E)=J_{+}(E)+J_{-}(E), \label{eq:J_eff_decomposition} \\ 
J_{\pm}(E)=\frac{1}{2} \frac{4t_h^2}{U\pm E}. \label{eq:Jpm} 
\end{gather}
The contributions $J_+$ and $J_-$ in Eq.~\eqref{eq:J_eff_decomposition} correspond to the perturbation processes (i) and (ii) shown in Fig.~\ref{fig:perturbation}~(a), respectively. These two contributions become inequivalent under the field. For simplicity, we focus on $E>0$. In this regime, as shown in Fig.~\ref{fig:perturbation}~(b), the dominant contribution is $J_{-}$ and its denominator decreases by $E$. This means that the energy cost decreases due to the tilted potential and the antiparallel spin configuration becomes more favored.

Let us move to the large tilt regime $|E|>U$. The most important point is that the derivation for the small tilt is directly applicable to this case. This is because the derivation in Appendix \ref{sec:static_derivation} formally uses the second-order perturbation for the \textit{singly occupied state} , and the physical origin of the singly occupied state is irrelevant to the derivation. As shown in Sec.~\ref{sec:charge_dynamics}, the doublon number in the Wannier-Stark regime has the same order of magnitude as in the Mott regime; thus, we can apply perturbation theory. 
In other words, the derivation is applicable to the entire regime with a small double occupancy, which includes both the Mott insulating and the Wannier-Stark localized regimes.
Therefore, the effective spin interaction in the large tilt case is also given by Eq.~\eqref{eq:J_eff}. 
In the large tilt case, the denominator of Eq.~\eqref{eq:J_eff} changes sign and the interaction becomes ferromagnetic. For $E>U$, the dominant contribution is $J_-$ and the corresponding energy cost $E-U$ becomes negative. Thus, the antiparallel spin configuration is energetically unfavorable. This is the origin of ferromagnetism.

\begin{figure*}
    \centering
    \includegraphics[width=15cm]{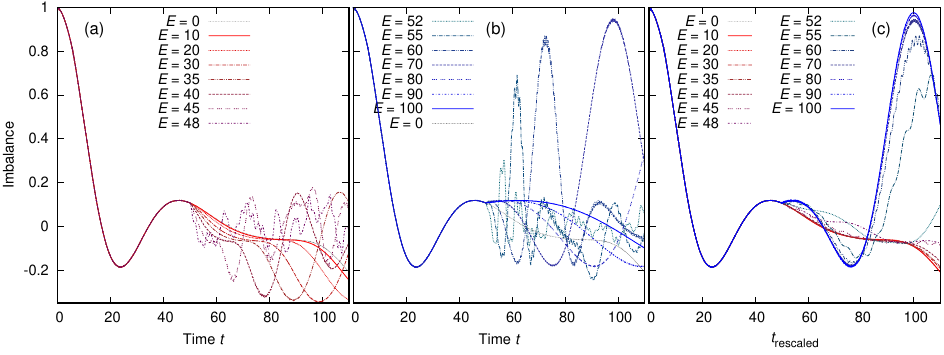}
    \caption{Time evolution of the local spin imbalance $\mathcal{I}_{L/2}$ [Eq.~\eqref{eq:imbalance}] for $L=10$ before and after the field $E$ is switched on at $t_\mathrm{on} = 50$. We set $U=50$ and $\Delta t = 1/3200$, and choose the singly occupied state $\vert\uparrow\downarrow\uparrow\downarrow\cdots\uparrow\downarrow\rangle$ as the initial state. We vary $E$ for $0\leq E\leq 2U$ and the data for $E \leq U$ and $E \geq U$ are shown in panels (a) and (b) respectively. Note that the data for $E=70$ in panel (b) exhibits almost perfect time-reversal dynamics because $J_\mathrm{eff}(E)/J_0$ ($\simeq -1.04$) is nearly minus one. In panel (c), we show all the data with a rescaled time defined in Eq.~\eqref{eq:rescaled_time}. $t_h$ ($t_h^{-1}$) is used as the unit of energy (time).
    }
    \label{fig:imbalance}
\end{figure*}

We also obtain a consistent effective model using Floquet theory. Applying the Floquet theory to the velocity-gauge Hamiltonian \eqref{eq:model_vgauge}, we employ the high-frequency (van Vleck) expansion~\cite{EckardtRMP2017} and obtain the effective Hamiltonian up to the order of $1/E^2$  as,
\begin{equation}
H_{\eff}^\mathrm{Floquet}\! =\! H_\mathrm{FM} + H_\mathrm{Hub}+H_\mathrm{pair}+H_b \label{eq:Heff_Floquet},
\end{equation}
where
\begin{align}
H_\mathrm{FM} &= -J_0  \left( \frac{U}{E} \right)^{\!\!2}~\!\!\sum_{i=1}^{L-1} \left(\bm S_i \cdot \bm S_{i+1} - \frac{n_i n_{i+1}}{4}\right),\\
H_\mathrm{Hub}&=U\left(1-\frac{4 t_h^2}{E^2}\right)\sum_{i=1}^L n_{i \uparrow} n_{i \downarrow},\\
H_\mathrm{pair}&= \frac{t_h^2 U}{E^2} \sum_{i=2}^{L-1} \sum_{\sigma=\uparrow,\downarrow} (c^\dagger_{i \sigma} c^\dagger_{i \bar{\sigma}} c_{i+1 \sigma} c_{i-1 \bar{\sigma}}+ \mathrm{h.c.}),\\ H_b &= \frac{t_h^2}{E} \sum_{\sigma=\uparrow,\downarrow} (n_{N \sigma} - n_{1\sigma}),
\end{align}
with $n_i = \sum_{\sigma=\uparrow,\downarrow} n_{i \sigma}$, $J_0 = 4t^2_h/U$, and $\bar{\uparrow}=\downarrow (\bar{\downarrow}=\uparrow)$. The detailed derivation is presented in Appendix~\ref{sec:Floquet_derivation}. Since the frequency is $E$, the high-frequency limit corresponds to the large tilt limit and thus the effective Hamiltonian~\eqref{eq:Heff_Floquet} becomes valid in the Wannier-Stark regime. Indeed, the hopping term vanishes in Eq.~\eqref{eq:Heff_Floquet} and it reflects the localization. While the Hamiltonian~\eqref{eq:Heff_Floquet} contains various interactions, such as the spin interaction $H_\mathrm{FM}$, the Hubbard interaction $H_\mathrm{Hub}$, and the pair hopping $H_\mathrm{pair}$, the most important part for us is that $H_\mathrm{FM}$ is ferromagnetic. Note that the coupling $-J_0 (U/E)^2$ is consistent with Eq.~\eqref{eq:J_eff} in the strong field limit $E/U \gg 1$. 

We comment on the related previous works and address the differences from them.
First, it is important to note that similar effective Hamiltonians for oscillating (AC) electric fields 
have been derived based on Floquet theory and their properties have been widely studied~\cite{Mentink2015, Bukov2016}. In contrast, the static (DC) case, which is studied in our study, has been less investigated. Several studies~\cite{Takasan2019, Furuya2021a, Katsura2009, Wang2014, Eckstein2017} have discussed effective spin interactions similar to Eq.~\eqref{eq:J_eff} for the DC electric field, but all these are limited to the Mott-insulating regime. Our work can be seen as an extension of them to the Wannier-Stark regime. Also, an effective model for the Wannier-Stark regime has been studied in the Bose-Hubbard model and it has already been pointed out that the ferromagnetic interaction is realized with a large tilt~\cite{Trotzky2008, Dimitrova2020}. The case of the fermionic Hubbard model has been studied very recently~\cite{Scherg2021, Kohlert2021}. While an effective model similar to our Floquet effective model~\eqref{eq:Heff_Floquet} has been derived in these papers, the Hamiltonian is not written in terms of spins and thus the appearance of ferromagnetism has not been apparent. In contrast, our spin Hamiltonian [Eqs.~\eqref{eq:H_eff_static} and~\eqref{eq:Heff_Floquet}] explicitly shows the ferromagnetic nature. Moreover, our work, as presented in the next section, shows the emergence of ferromagnetism through the real-time dynamics of the Hubbard model, which is independent of the effective Hamiltonian.

\section{Real-time spin dynamics}
\label{sec:real-time_dynamics}
In the previous section, we have derived the effective model under a tilt and pointed out the emergence of ferromagnetism. However, the discussion has been based on perturbation theory and it is still unclear whether the result is robust beyond the approximation. Also, it is not yet clear how to find the signature of ferromagnetism in the observable quantities. Naively, the magnetization induced by a tilt might be regarded as clear evidence, but it is difficult to observe when we start from the untilted Mott insulator. This is because the time evolution with Eq.~\eqref{eq:model_lgauge} (equivalent to Eq.~\eqref{eq:model_vgauge}) conserves the total magnetization and the original Mott insulator has zero magnetization in the low-temperature state. 

To clarify these points, we study the real-time spin dynamics and show how to extract information about the effective interaction. As in Sec.~\ref{sec:charge_dynamics}, we solve the many-body Schr\"odinger equation with the Hamiltonian Eq.~\eqref{eq:model_vgauge} using the fourth-order Runge--Kutta method~\footnote{The numerical method is explained in Appendix~\ref{subsec:RK4}.} and obtain the time evolution starting from the Ne\'el state $\ket{\uparrow \downarrow \uparrow \downarrow \cdots}$. To see the spin dynamics, we study the local spin imbalance 
\begin{gather}
\mathcal{I}_j = \langle\Psi(t)\vert n_{ j \uparrow} - n_{j \downarrow} \vert\Psi(t)\rangle. \label{eq:imbalance}
\end{gather}
To suppress the boundary effect, we focus on $j=\lfloor L/2 \rfloor$ in our numerical calculation~\footnote{While the boundary effect becomes larger near the system edges, the qualitative behaviors do not depend much on the position. For details, see Appendix \ref{subsec:boundary}.}. In order to see the sharp contrast between $E=0$ and $E\neq0$, we first consider the time evolution without a tilt from $t=0$ to $t=t_\mathrm{on}(=50)$ and then apply the electric field after $t=t_\mathrm{on}$. 

To clarify the effect of the exchange interaction on the real-time dynamics, we consider the time-evolution operator with the effective Hamiltonian \eqref{eq:H_eff_static} $U_\eff(t)=e^{-i H_\eff t}=\exp[-i f(E)t \sum_{j=1}^L J_0 \bm S_j \cdot \bm S_{j+1}]$ where $J_\eff(E)=f(E)J_0$ and 
\begin{equation}f(E)=[1-(E/U)^2]^{-1}.\label{eqn:fE}
\end{equation}
Under the strong coupling condition $U \gg t_h$, the dynamics is expected to be governed by this operator~\footnote{The dynamics for smaller $U$ are presented in Appendix \textbf{\ref{sec:smallerU}}. While the dynamics deviates from the one for the effective Hamiltonian with approaching the resonant point $|E| \sim U$, the signature in the spin imbalance survives until around $U\simeq 10$.}. The remarkable feature is that the $E$-dependence only appears as $f(E)t$ and $f(E)$ works as the scale factor in the time direction. This means that the time evolution is accelerated in the Mott-insulating regime where $J_\eff$ is enhanced. In contrast, the time evolution is reversed in the Wannier-Stark regime since $J_\eff$ becomes negative. Indeed, these features are seen in the time evolution of the local spin imbalance shown in Fig.~\ref{fig:imbalance}~(a) and (b). 
To clearly see whether the dynamics follows the effective Hamiltonian, we show these data with a rescaled time $t_\mathrm{rescaled}$, defined as 
\begin{align}
    t_\mathrm{rescaled} = 
    \begin{cases}
    t & (t < t_\mathrm{on}), \\
    t_\mathrm{on} + |f(E)| (t-t_\mathrm{on}) & (t > t_\mathrm{on}),
    \end{cases} \label{eq:rescaled_time}
\end{align}
in Fig.~\ref{fig:imbalance}~(c). As seen in this figure, the data collapse into two curves except for the resonant regime $U \simeq E$. This means that the spin dynamics is well-described by the effective Hamiltonian~\eqref{eq:H_eff_static} in both the Mott-insulating and the Wannier-Stark regime. It shows that the exchange coupling under a tilt is given by Eq.~\eqref{eq:J_eff} and thus ferromagnetism is realized with a large tilt. We emphasize that these results are obtained only from the Hubbard model and do not depend on any specific approximation, such as  perturbation theory. This is the most important result in this paper. Note that a similar time evolution of spins in the Hubbard model under oscillating electric fields was already studied in Ref.~\cite{Mentink2015}.

\begin{figure}
    \centering
    \includegraphics[width=7cm]{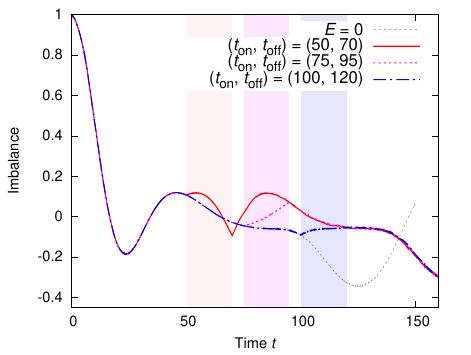}
    \caption{Time evolution of the local spin imbalance $\mathcal{I}_{L/2}$ [Eq.~\eqref{eq:imbalance}] for $L=10$ when the field $E = 70.71 \sim \sqrt{2}U$ is turned on at $t_\mathrm{on}$ and switched off at $t_\mathrm{off} = t_\mathrm{on} + \delta t$ with $\delta t = 20$. We set $U=50$ and $\Delta t = 1/3200$, and choose the singly occupied state $\vert\uparrow\downarrow\uparrow\downarrow\cdots\uparrow\downarrow\rangle$ as the initial state.
    We vary $t_\mathrm{on}$ for $t_\mathrm{on}=50, 75, 100$ and they are shown with different line types. The time duration when $E$ is finite are shown with the shaded regions respectively. The dynamics without $E$ is shown in a gray dashed curve. $t_h$ ($t_h^{-1}$) is used as the unit of energy (time).
    }
    \label{fig:on-off}
\end{figure}

An interesting application of this dynamics is the realization of time reversal. Since $f(E)$ takes $-1$ at $E=\sqrt{2}U$, the time evolution is exactly reversed. In Fig.~\ref{fig:on-off}, we plot the time evolution when the field $E~(=70.71\sim \sqrt{2}U)$ is turned on at various $t_\mathrm{on}~(=50, 75, 100)$ and switched off at $t_\mathrm{off} = t_\mathrm{on} + \delta t$ with $\delta t = 20$. The spin dynamics is obviously reversed for the duration of $\delta t$ in Fig.~\ref{fig:on-off}. Also, all the data almost coincide after $t=120$ in Fig.~\ref{fig:on-off}. This demonstrates the accuracy of the time reversal. The time-reversal dynamics is known to be useful for experimentally measuring out-of-time-ordered correlators (OTOCs), which are used as diagnostics for chaos in quantum systems~\cite{Larkin1969, Maldacena2016}. While there have been many efforts to realize time reversal~\cite{Li2017, Garttner2017}, it is still not an easy task. Our findings suggest that time reversal in the strongly correlated Hubbard model is achieved simply by adding a tilt. Since the controls of tilt have already been achieved in various AMO systems~\cite{Simon2011, GuardadoSanchez2020, Guo2020, Dimitrova2020, Scherg2021, Kohlert2021, Morong2021}, our protocol can be useful for experimentally measuring the OTOCs.

\section{Experimental platforms}
We discuss the experimental platforms for observing ferromagnetism in tilted Mott insulators. For this purpose, AMO systems are promising because many-body quantum phenomena induced by a large tilted potential have already been studied experimentally in various AMO systems, such as ultracold atoms~\cite{Meinert2014, Geiger2018, Chen2011, Simon2011, GuardadoSanchez2020, Scherg2021, Kohlert2021, Trotzky2008, Dimitrova2020}, trapped ions~\cite{Morong2021}, and superconducting qubits~\cite{Guo2020}. In particular, ultracold fermionic atoms in an optical lattice~\cite{GuardadoSanchez2020, Scherg2021, Kohlert2021} provide an ideal platform for the fermionic Hubbard model \eqref{eq:model_lgauge} and thus they are the most promising platform for our study. The spin-resolved dynamics can be obtained in this setup and thus the imbalance dynamics shown in Fig.~\ref{fig:imbalance} will be directly observable. In solid-state electronic systems, it is difficult to realize a large electric field that can achieve the Wannier-Stark regime~\cite{Schmidt2018}. To avoid this difficulty, synthetic structured systems, such as semiconductor superlattices~\cite{Mendez1988, Voisin1988}, have been used. The ferromagnetism in the Wannier-Stark regime can be observed in such systems. For this purpose, an array of quantum dots is a good candidate because the fermionic Hubbard model~\cite{Hensgens2017} and the effective Heisenberg spin chain~\cite{vanDiepen2021} become possible to be simulated in this setup thanks to the recent developments in experimental techniques. A more challenging direction is the observation in bulk solids. The transient signature of the Wannier-Stark ladder is observed in a bulk semiconductor~\cite{Schmidt2018}, and thus similar pump-probe type measurements in strongly correlated materials may enable us to observe the ferromagnetic signature in Mott insulator materials.

\section{Summary and Outlook}

In this paper, we have studied the spin interaction in fermionic Mott insulators with a tilt. Using perturbation theory and direct calculation of the real-time evolution, we have revealed the effective interaction in both the small and large tilt regimes and found the emergence of ferromagnetism. This appears in the change of speed and time direction in the real-time dynamics, which can be observed in various experimental platforms.

Finally, we address several open issues to be studied in the future. One is the extension of our effective spin models to more general settings. Since our derivation is a very simple second-order perturbation theory, it is easy to apply it to various Hubbard-type models. Indeed, previous studies have studied multi-orbital or spin-orbit-coupled Hubbard models and provided the Heisenberg interaction for higher spins (e.g., $S=1$~\cite{Takasan2019}), the superexchange interaction~\cite{Furuya2021a}, and the Dzyaloshinskii–Moriya interaction~\cite{Furuya2021b} under a small tilted potential that preserves the Mott-insulating states. Thus, it is promising to study them in the Wannier-Stark regime and discuss their physical implications. Also, our derivation of the effective model is applicable to arbitrary dimensions, although we have only treated a one-dimensional model for simplicity. The effect of tilt on magnetic orders and topological phases in higher dimensions is also an interesting direction as it has been examined in the Mott-insulating regime~\cite{Takasan2019, Furuya2021a, Furuya2021b}. Ultracold atom experiments with a tilted optical lattice in two dimensions have already been performed~\cite{Adler2024} and thus the study along this direction will be relevant for experiments.

Another direction is to study the relation to quantum phase transitions triggered by ferromagnetic interactions. While we have studied the emergence of ferromagnetic interaction, we have not considered the phase transition driven by it. As mentioned at the beginning of Sec.~\ref{sec:real-time_dynamics}, the real-time dynamics studied in this paper conserves the total spin, and it is difficult to induce ferromagnetic order starting from the antiferromagnetic Mott insulating state. One way to overcome this difficulty is to add a small dissipation that allows the system to become magnetized. A similar setup in the Floquet prethermalization in open quantum systems has already been studied and exhibits a finite magnetization~\cite{Ikeda2020}. Thus, we expect that our setup can be studied in a similar manner. One remark is that dissipation can break the Wannier-Stark localization in the long-time limit~\cite{Wu2019}, and thus we need to find a certain intermediate time scale. If we limit ourselves to closed quantum systems, one way to capture the signature of quantum phase transitions is to study so-called dynamical quantum phase transitions (DQPTs)~\cite{Heyl2018}, where the dynamical free energy defined with time-dependent wavefunctions exhibits singularities in time. As the DQPT has been observed in the interaction quench in the Ising model~\cite{Heyl2018}, it might be possible to observe a similar DQPT in our model by changing the tilt. Since the DQPT has already been observed in experiments with ultracold atoms~\cite{Flaschner2018} and trapped ions~\cite{Jurcevic2017}, this direction is important for finding an ideal experimental platform.

\begin{acknowledgments}
We are thankful to Ehud Altman, Marin Bukov, Hosho Katsura, Norio Kawakami, Kensuke Kobayashi, Kaoru Mizuta, Joel E. Moore, Takashi Oka, and Masafumi Udagawa for their valuable discussions. K.T. thanks Masahiro Sato for the previous collaborations that are closely related to this project. K.T. was supported by the U.S. Department of Energy (DOE), Office of Science, Basic Energy Sciences (BES), under Contract No. AC02-05CH11231 within the Ultrafast Materials Science Program (KC2203), by JSPS KAKENHI
(Grant Nos. JP22K20350, JP23K17664, and JP25K17312), and by JST PRESTO (Grant Nos. JPMJPR2256 and JPMJPR2596). M.T. was supported by JSPS KAKENHI (KAKENHI Grants Nos. JP17K17822, JP20K03787, JP20H05270, JP21H05185, and JP25K00925) and JST CREST (Grant No. JPMJCR24I2).
\end{acknowledgments}

\appendix

\section{Static perturbation theory}
\label{sec:static_derivation}
In this section, we derive the effective model \eqref{eq:H_eff_static}. This derivation is formally the same as in Ref.~\cite{Takasan2019}, where only the small tilt case was discussed.
First, we start with a generic Hubbard model where we do not specify the form of the on-site potential and the lattice structure. After deriving the effective model, we will apply the result to our model \eqref{eq:model_lgauge} in the main text. Let us consider the fermionic Hubbard model with an arbitrary on-site potential term, 
\begin{align}
H &= H_t + H_U + H_V, \label{Model_Onsite}
\end{align}
with
\begin{align}
H_t &= \sum_{ij} \sum_{\sigma=\uparrow, \downarrow}  t_{ij} c^\dagger_{i \sigma} c_{j \sigma} \\
H_U &= U \sum_i n_{i \uparrow} n_{i \downarrow} \\
H_V &= \sum_{i} \sum_{\sigma=\uparrow, \downarrow} V_i n_{i \sigma}, 
\end{align}
where the indices $i$ and $j$ run over all the sites. Here, we consider repulsive interaction $U>0$ and the half-filled case. 

To derive the effective spin Hamiltonian, we introduce a projection operator $P_g$ which projects states into the singly-occupied subspace. We also define the complementary projection operator $P_e \equiv \bm 1 - P_g$ where $\bm 1$ is the identity operator on the total Hilbert space. Using these operators and the second order perturbation theory, we can write down the effective Hamiltonian up to the second order correction as
\begin{align}
H_\eff = H_{gg} + H_{ge} \frac{1}{E_g - H_{ee}} H_{eg}, 
\label{Formula_Heff}
\end{align}
where $H_{\alpha \beta} = P_\alpha H P_\beta$ $(\alpha, \beta = g, e)$ and 
$E_g$ is defined by $H_{gg} \ket{\Psi_g} = E_g \ket{\Psi_g}$. We apply this formula (\ref{Formula_Heff}) to the Hubbard model (\ref{Model_Onsite}). This perturbation theory is usually applied to the Mott-insulating state because the singly-occupied states form the ground state subspace and the effective Hamiltonian gives the important information about the low-temperature physics. However, we do not have to specify the physical origin of the singly-occupied state when we apply the perturbation theory. If a state close to the singly-occupied state is realized, this effective Hamiltonian gives a reasonable description, regardless of its origin. This is the reason why we can apply the same effective model to the Wannier-Stark localized regime.

Let us consider $H_{eg}$. Among the three terms $H_t$, $H_U$ and $H_V$, only the hopping $H_t$ has 
a matrix element between the singly occupied states and the other states.
Therefore, $H_{eg}$ is written as 
\begin{align}
H_{eg}&=P_e \left( \sum_{ij} \sum_{\sigma=\uparrow,\downarrow}  t_{ij} c^\dagger_{i \sigma} c_{j \sigma} \right) P_g.
\end{align}
Considering the exclusion principle, $H_{eg}$ survives 
only when the spin indices $\sigma$ on the $i$-th and $j$-th sites are different. Thus, we can rewrite $H_{eg}$ as  
\begin{align}
H_{eg}&= \sum_{ij} \sum_{\sigma=\uparrow,\downarrow}  t_{ij} c^\dagger_{i \sigma} c_{j \sigma} (S^z_i - S^z_j)^2.
\end{align}

Next we compute the energy difference between the singly-occupied state and the intermediate states which are shown in Fig.~\ref{fig:perturbation}~(a) in the main text. Let us focus on the two sites ($i$-th site and $j$-th site) and a hopping process from the $j$-th site to the $i$-th site. In the singly occupied states, both sites have a single particle respectively and thus the energy contribution from these sites is $V_i + V_j$. In contrast, the $i$-th site is doubly occupied and the $j$-th site is vacant in the intermediate states. Thus, the energy is $U + 2V_i$. Summing up all the contributions, we obtain
\begin{align}
&\frac{1}{E_g - \calH_{ee}} \calH_{eg}\nonumber\\ &=  
\sum_{ij}\sum_{\sigma=\uparrow,\downarrow} \frac{1}{(V_i + V_j) - (U + 2V_i)}  t_{ij} c^\dagger_{i \sigma} c_{j \sigma} (S^z_i - S^z_j)^2 \nonumber \\
&= -\sum_{ij }\sum_{\sigma=\uparrow,\downarrow}\frac{1}{U - \Delta V_{ij}}   t_{ij} c^\dagger_{i \sigma} c_{j \sigma} (S^z_i - S^z_j)^2, \label{eq:inter}
\end{align}
with $\Delta V_{ij} = V_i  -V_j$.

Finally, we operate the $\calH_{ge}$ to Eq.~(\ref{eq:inter}) and 
then the second-order perturbation term is calculated as follows: 
\begin{align}
&\calH_{ge} \frac{1}{E_g - \calH_{ee}} \calH_{eg}\\
&= - P_g \sum_{i^\prime j^\prime}\sum_{\sigma^\prime=\uparrow,\downarrow}  t_{j^\prime i^\prime} c^\dagger_{j^\prime \sigma^\prime} c_{i^\prime \sigma^\prime} \times \nonumber \\
&\qquad \qquad \sum_{ij} \sum_{\sigma=\uparrow,\downarrow}\frac{1}{U - \Delta V_{ij}}   t_{ij} c^\dagger_{i \sigma} c_{j \sigma} (S^z_i - S^z_j)^2 \nonumber\\
&= - \sum_{ij}\sum_{\sigma=\uparrow,\downarrow} \frac{|t_{ij}|^2}{U - \Delta V_{ij}} c^\dagger_{j \sigma} c_{i \sigma}  c^\dagger_{i \sigma} c_{j \sigma} (S^z_i - S^z_j)^2 \nonumber \\  
&\qquad \qquad  - \sum_{ij}\sum_{\sigma=\uparrow,\downarrow} \frac{|t_{ij}|^2}{U - \Delta V_{ij}} c^\dagger_{j \bar{\sigma}} c_{i \bar{\sigma}}  c^\dagger_{i \sigma} c_{j \sigma} (S^z_i - S^z_j)^2  \nonumber\\
&= - \sum_{ij} \frac{|t_{ij}|^2}{U - \Delta V_{ij}} (S^z_i - S^z_j)^2 \nonumber \\ &\qquad + \sum_{ij} \frac{|t_{ij}|^2}{U - \Delta V_{ij}} (S^-_j S^+_i + S^+_j S^-_i) \nonumber\\
&= \sum_{ij} \frac{2 |t_{ij}|^2}{U - \Delta V_{ij}} \bm S_i \cdot \bm S_j + \mathrm{const.}, \label{eq:final}
\end{align}
where we have defined $\bar{\sigma} = - \sigma$. Since $\calH_{gg}$ only gives a constant term, the effective Hamiltonian up to the second-order is
\begin{align}
H_\eff &= \sum_{ij} \frac{2 |t_{ij}|^2}{U - \Delta V_{ij}} \bm S_i \cdot \bm S_j+ \mathrm{const.}\nonumber\\ 
&= \sum_{\langle ij \rangle} \frac{4 |t_{ij}|^2}{U} \frac{1}{1- \left( \frac{\Delta V_{ij}}{U} \right)^2} \bm S_i \cdot \bm S_j+ \mathrm{const.},
\end{align}
where the summation is taken over the every pair $\langle i, j \rangle$ in the last line. When we apply this result to our model \eqref{eq:model_lgauge} in the main text, we reach the effective model \eqref{eq:H_eff_static} in the main text.

\section{Floquet perturbation theory}
\label{sec:Floquet_derivation}
In this section, we derive the effective model \eqref{eq:Heff_Floquet}  using the Floquet theory.

\subsection{Effective Hamiltonian and high-frequency expansion in Floquet theory}
We start with the velocity-gauge Hamiltonian \eqref{eq:model_vgauge}. Since this model is time-periodic with the period $2 \pi/ E$, we can apply the Floquet theory, which is a theoretical framework for time-periodic systems~\cite{EckardtRMP2017}. In the Floquet theory, the effective static Hamiltonian in the high-frequency limit plays an important role and the methods to calculate it have been established~\cite{EckardtRMP2017}. Since the frequency is $E$ in the model \eqref{eq:model_vgauge}, the high-frequency expansion is expected to work in a large tilt (i.e., Wannier-Stark) regime. Thus, we apply them to the model \eqref{eq:model_vgauge} and derive the effective model for the Wannier-Stark regime.

To introduce the effective Hamiltonian in the Floquet theory, we define the time-evolution operator $U(t, t_0)=\mathcal{T}\exp[-i \int^t_{t_0} dt H(t)]$ and consider its decomposition as $U(t, t_0) = e^{-i K (t)}e^{-i H_\eff (t-t_0)} e^{i K (t_0)}$.
Here, $H_\eff$ is a static operator called an effective Hamiltonian and $K (t)$ is a time-periodic operator called a kick operator. The operators $H_\eff$ and $K(t)$ are known to be perturbatively expanded in the power of $1/\omega$~\cite{EckardtRMP2017}. Using this expansion, the effective Hamiltonian is given as 
\begin{align}
    H^{(n_0)}_{\eff} &= \sum^{n_0}_{n=0} H_{\eff, n} \\
    H_{\eff, 0} &= H_0 \\
    H_{\eff, 1} &= \sum^\infty_{m=1} \frac{[H_m, H_{-m}]}{m \omega} \\
    H_{\eff, 2} &= \sum_{m \neq 0} \frac{[[H_m, H_0], H_{-m}]}{2m^2 \omega^2} \nonumber \\
    &\qquad + \sum_{m \neq 0} \sum_{m^\prime \neq 0, m} \frac{[[H_m, H_{m^\prime-m}], H_{-m^\prime}]}{3mm^\prime \omega^2},
\end{align}
where $H_n = \frac{1}{T}\int^{T/2}_{-T/2}dt H(t) e^{- in \omega t}$. $n_0$ is the truncation order and we consider $n_0 = 2$ in this study. This expansion is called the van Vleck expansion~\cite{Mikami2016}. For latter convenience, we summarize the Fourier modes of the model (\ref{eq:model_vgauge}) as
\begin{align}
H_0&= U \sum_{i=1}^L n_{i \uparrow} n_{i \downarrow},\\
H_{+1} &= -t_h \sum_{i=1}^{L-1} \sum_{\sigma=\uparrow,\downarrow} c^\dagger_{i+1 \sigma} c_{i \sigma}, \\
H_{-1} &= -t_h \sum_{i=1}^{L-1} \sum_{\sigma=\uparrow,\downarrow} c^\dagger_{i \sigma} c_{i+1 \sigma},  
\end{align}
and $H_{|n|>1} = 0$. Using these terms, we calculate the effective Hamiltonian order by order.
\subsection{Zero-th order}
The zero-th order contribution is given as
\begin{align}
H_{\eff, 0} = H_0 = U \sum_{i=1}^{L} n_{i \uparrow} n_{i \downarrow}. \label{eq:Heff_Floquet_expansion_0}   
\end{align}

\subsection{First order}
The first order term is $H_{\eff, 1}=\sum_n [H_{+n} , H_{-n}]/(n E)=[H_{+1} , H_{-1} ]/E$ and this commutator is computed as
\begin{align}
[H_{+1} , H_{-1} ] 
&= t_h^2  \sum_{i, j=1}^{L-1} \sum_{\sigma, \sigma^\prime = \uparrow, \downarrow}  [c^\dagger_{i+1 \sigma} c_{i \sigma}, c^\dagger_{j \sigma^\prime} c_{j+1 \sigma^\prime}] \nonumber \\
&=  t_h^2 \sum_{i=1}^{L-1} \sum_{\sigma=\uparrow, \downarrow} c^\dagger_{i+1 \sigma} c_{i+1 \sigma} - t_h^2 \sum_{i=1}^{L-1} \sum_{\sigma=\uparrow, \downarrow} c^\dagger_{i \sigma} c_{i \sigma} \nonumber \\
&= t_h^2 \sum_{\sigma=\uparrow, \downarrow}(n_{N \sigma} - n_{1\sigma}).
\end{align}
Therefore,
\begin{align}
H_{\eff, 1} &= \frac{t_h^2}{E} \sum_{\sigma=\uparrow, \downarrow} (n_{N \sigma} - n_{1\sigma}) \label{eq:Heff_Floquet_expansion_1}
\end{align}
Note that this term vanishes if we adopt the periodic boundary condition.

\begin{widetext}

\subsection{Second order}
Since the higher-order Fourier components ($n>1$) are zero, the second-order term is simplified as
\begin{align}
H_{\eff, 2}  &= \frac{[[H_{+1},H_{0}],H_{-1}]}{2 E^2} + \frac{[[H_{-1},H_{0}],H_{+1}]}{2 E^2}.
\end{align}
We calculate the nested commutators $[[H_{+1},H_{0}],H_{-1}]$ and $[[H_{-1},H_{0}],H_{+1}]$. First, we obtain
\begin{align}
[H_{+1},H_{0}] 
&= -t_h U \sum_{i=1}^{L-1} \left( 
c^\dagger_{i+1 \uparrow} c_{i \uparrow} c^\dagger_{i \downarrow} c_{i \downarrow} 
- c^\dagger_{i+1 \uparrow} c_{i \uparrow} c^\dagger_{i+1 \downarrow} c_{i+1 \downarrow}  
+ c^\dagger_{i \uparrow} c_{i \uparrow} c^\dagger_{i+1 \downarrow} c_{i \downarrow} 
- c^\dagger_{i+1 \uparrow} c_{i+1 \uparrow} c^\dagger_{i+1 \downarrow} c_{i \downarrow} \right ), \\
[H_{-1},H_{0}] 
&= -t_h U \sum_{i=1}^{L-1}
\left(
c^\dagger_{i \uparrow} c_{i+1 \uparrow} c^\dagger_{i+1 \downarrow} c_{i+1 \downarrow} 
- c^\dagger_{i \uparrow} c_{i+1 \uparrow} c^\dagger_{i \downarrow} c_{i \downarrow}
+c^\dagger_{i+1 \uparrow} c_{i+1 \uparrow} c^\dagger_{i \downarrow} c_{i+1 \downarrow} 
- c^\dagger_{i \uparrow} c_{i \uparrow} c^\dagger_{i \downarrow} c_{i+1 \downarrow} \right ).
\end{align}
Then, we calculate the nested ones as 
\begin{align}
[[H_{+1},H_{0}],H_{-1}]
&= t_h^2 U \sum_{i=1}^{L-1} \left(c^\dagger_{i+1 \uparrow} c_{i+1 \uparrow} c^\dagger_{i \downarrow} c_{i \downarrow} - c^\dagger_{i \uparrow} c_{i \uparrow} c^\dagger_{i \downarrow} c_{i \downarrow} - c^\dagger_{i+1 \uparrow} c_{i+1 \uparrow} c^\dagger_{i+1 \downarrow} c_{i+1 \downarrow} + c^\dagger_{i \uparrow} c_{i \uparrow} c^\dagger_{i+1 \downarrow} c_{i+1 \downarrow} \right.\nonumber\\ &\qquad \qquad + c^\dagger_{i \uparrow} c_{i+1 \uparrow} c^\dagger_{i+1 \downarrow} c_{i \downarrow} - c^\dagger_{i-1 \uparrow} c_{i \uparrow} c^\dagger_{i+1 \downarrow} c_{i \downarrow} - c^\dagger_{i+1 \uparrow} c_{i+2 \uparrow} c^\dagger_{i+1 \downarrow} c_{i \downarrow} + c^\dagger_{i \uparrow} c_{i+1 \uparrow} c^\dagger_{i+1 \downarrow} c_{i \downarrow}\nonumber\\ &\qquad \qquad \quad + c^\dagger_{i+1 \uparrow} c_{i \uparrow} c^\dagger_{i \downarrow} c_{i+1 \downarrow} - c^\dagger_{i+1 \uparrow} c_{i \uparrow} c^\dagger_{i-1 \downarrow} c_{i \downarrow} - c^\dagger_{i+1 \uparrow} c_{i \uparrow} c^\dagger_{i+1 \downarrow} c_{i+2 \downarrow} + c^\dagger_{i+1 \uparrow} c_{i \uparrow} c^\dagger_{i \downarrow} c_{i+1 \downarrow}\nonumber\\  &\left. \qquad \qquad \qquad + c^\dagger_{i \uparrow} c_{i \uparrow} c^\dagger_{i+1 \downarrow} c_{i+1 \downarrow} - c^\dagger_{i \uparrow} c_{i \uparrow} c^\dagger_{i \downarrow} c_{i \downarrow} - c^\dagger_{i+1 \uparrow} c_{i+1 \uparrow} c^\dagger_{i+1 \downarrow} c_{i+1 \downarrow} + c^\dagger_{i+1 \uparrow} c_{i+1 \uparrow} c^\dagger_{i \downarrow} c_{i \downarrow} \right),
\end{align}
\begin{align}
[[H_{-1},H_{0}],H_{+1}]
&=t_h^2 U \sum_{i=1}^{L-1}\left(
 c^\dagger_{i \uparrow} c_{i \uparrow} c^\dagger_{i+1 \downarrow} c_{i+1 \downarrow} - c^\dagger_{i+1 \uparrow} c_{i+1 \uparrow} c^\dagger_{i+1 \downarrow} c_{i+1 \downarrow} - c^\dagger_{i \uparrow} c_{i \uparrow} c^\dagger_{i \downarrow} c_{i \downarrow} + c^\dagger_{i+1 \uparrow} c_{i+1 \uparrow} c^\dagger_{i \downarrow} c_{i \downarrow} \right.\nonumber\\ & \qquad \qquad + c^\dagger_{i+1 \uparrow} c_{i \uparrow} c^\dagger_{i \downarrow} c_{i+1 \downarrow} - c^\dagger_{i+2 \uparrow} c_{i+1 \uparrow} c^\dagger_{i \downarrow} c_{i+1 \downarrow} - c^\dagger_{i \uparrow} c_{i-1 \uparrow} c^\dagger_{i \downarrow} c_{i+1 \downarrow} + c^\dagger_{i+1 \uparrow} c_{i \uparrow} c^\dagger_{i \downarrow} c_{i+1 \downarrow}\nonumber\\ &\qquad \qquad \quad + c^\dagger_{i \uparrow} c_{i+1 \uparrow} c^\dagger_{i+1 \downarrow} c_{i \downarrow} - c^\dagger_{i \uparrow} c_{i+1 \uparrow} c^\dagger_{i+2 \downarrow} c_{i+1 \downarrow} - c^\dagger_{i \uparrow} c_{i+1 \uparrow} c^\dagger_{i \downarrow} c_{i-1 \downarrow} + c^\dagger_{i \uparrow} c_{i+1 \uparrow} c^\dagger_{i+1 \downarrow} c_{i \downarrow}\nonumber\\ &\left. \qquad \qquad \qquad + c^\dagger_{i+1 \uparrow} c_{i+1 \uparrow} c^\dagger_{i \downarrow} c_{i \downarrow} - c^\dagger_{i+1 \uparrow} c_{i+1 \uparrow} c^\dagger_{i+1 \downarrow} c_{i+1 \downarrow} - c^\dagger_{i \uparrow} c_{i \uparrow} c^\dagger_{i \downarrow} c_{i \downarrow} + c^\dagger_{i \uparrow} c_{i \uparrow} c^\dagger_{i+1 \downarrow} c_{i+1 \downarrow} \right) .
\end{align}
Therefore,
\begin{align}
H_{\eff, 2}
&= - \frac{4 t_h^2 U}{E^2} \sum_{i=1}^{L} n_{i \uparrow} n_{i \downarrow}
+ 2 t_h^2 U \sum_{i=1}^{L-1}
\left(
n_{i+1 \uparrow} n_{i \downarrow} +n_{i \uparrow} n_{i+1 \downarrow}
\right) + \frac{2 t_h^2 U}{E^2} \sum_{i=1}^{L-1}
\left(
c^\dagger_{i \uparrow} c_{i+1 \uparrow} c^\dagger_{i+1 \downarrow} c_{i \downarrow}+
c^\dagger_{i+1 \uparrow} c_{i \uparrow} c^\dagger_{i \downarrow} c_{i+1 \downarrow}
\right ) \nonumber \\
&\qquad- \frac{t_h^2 U}{E^2}\sum_{i=2}^{L-1} \left( 
c^\dagger_{i \uparrow} c_{i+1 \uparrow} c^\dagger_{i \downarrow} c_{i-1 \downarrow}+
c^\dagger_{i-1 \uparrow} c_{i \uparrow} c^\dagger_{i+1 \downarrow} c_{i \downarrow}+
c^\dagger_{i \uparrow} c_{i-1 \uparrow} c^\dagger_{i \downarrow} c_{i+1 \downarrow}+
c^\dagger_{i+1 \uparrow} c_{i \uparrow} c^\dagger_{i-1 \downarrow} c_{i \downarrow}
\right ). \label{eq:Heff_Floquet_expansion_2}
\end{align}

\subsection{Effective Hamiltonian $H_\eff^{(2)}$}

Using the above results \eqref{eq:Heff_Floquet_expansion_0}, \eqref{eq:Heff_Floquet_expansion_1}, and \eqref{eq:Heff_Floquet_expansion_2}, we obtain the effective Hamiltonian up to the second order as
\begin{align}
H_{\eff}^{(2)} &=  H_\mathrm{int1}  + H_\mathrm{int2} + H_\mathrm{Hub} + H_\mathrm{pair}+H_b,
\end{align}
with
\begin{align}
H_\mathrm{int1} &= 
\frac{2 t^2 U}{E^2} \sum_{i=1}^{L-1} \left(
n_{i+1 \uparrow} n_{i \downarrow} + n_{i \uparrow} n_{i+1 \downarrow}
\right), \\
H_\mathrm{int2}&=
-\frac{2 t^2 U}{E^2} \sum_{i=1}^{L-1}
\left(
c^\dagger_{i \uparrow}c_{i \downarrow} c^\dagger_{i+1 \downarrow} c_{i+1 \uparrow} +
 c^\dagger_{i \downarrow} c_{i \uparrow}c^\dagger_{i+1 \uparrow} c_{i+1 \downarrow}
\right ), \\
H_\mathrm{Hub}&=U \left ( 1 - \frac{4t_h^2}{E^2}  \right) \sum_{i=1}^{L} n_{i \uparrow} n_{i \downarrow},
\\
H_\mathrm{pair}&= 
-\frac{t^2 U}{E^2} 
\sum_{i=2}^{L-1}
\left(
c^\dagger_{i \uparrow} c_{i+1 \uparrow} c^\dagger_{i \downarrow} c_{i-1 \downarrow}+
c^\dagger_{i \uparrow} c_{i-1 \uparrow} c^\dagger_{i \downarrow} c_{i+1 \downarrow}+
\mathrm{h.c.}\right ), \\
H_{b} &= \frac{t_h^2}{E} \sum_{\sigma=\uparrow, \downarrow} (n_{N \sigma} - n_{1\sigma})
\end{align}

The sum of $H_\mathrm{int1}$ and $H_\mathrm{int2}$ is rewritten with the spin operator $\bm S_i$. Using the following relation
\begin{align}
\bm S_i \cdot \bm S_j - \frac{n_i n_j}{4} &= \frac{1}{2} (S^+_i S^-_j + S^-_i S^+_j) + S^z_i S^z_j  - \frac{n_i n_j}{4} \nonumber
\\
&= \frac{1}{2} \left( c^\dagger_{i \uparrow}c_{i \downarrow} c^\dagger_{j \downarrow} c_{j\uparrow} +c^\dagger_{i \downarrow} c_{i \uparrow} c^\dagger_{j\uparrow} c_{j \downarrow} - n_{i \uparrow} n_{j \downarrow} - n_{i \downarrow} n_{j \uparrow} \right),
\end{align}
the sum is written as
\begin{align}
H_\mathrm{int1}+H_\mathrm{int2}&=
\frac{2 t^2 U}{E^2} \sum_{i=1}^{L-1}
\left( n_{i+1 \uparrow} n_{i \downarrow} + n_{i+1 \downarrow} n_{i \uparrow}
-c^\dagger_{i \uparrow}c_{i \downarrow} c^\dagger_{i+1 \downarrow} c_{i+1 \uparrow}
-c^\dagger_{i \downarrow} c_{i \uparrow}c^\dagger_{i+1 \uparrow} c_{i+1 \downarrow}
\right )\nonumber \\
&=-\frac{4 t^2 U}{E^2} \sum_{i=1}^{L-1} \left( \bm S_i \cdot \bm S_{i+1} -\frac{n_i n_j}{4} \right) \equiv H_\mathrm{FM}.
\end{align}
We arrive at the form of the effective Hamiltonian shown in Eq.~(\ref{eq:Heff_Floquet}) in the main text.

\end{widetext}

\begin{figure*}
    \centering
    \includegraphics[width=15cm]{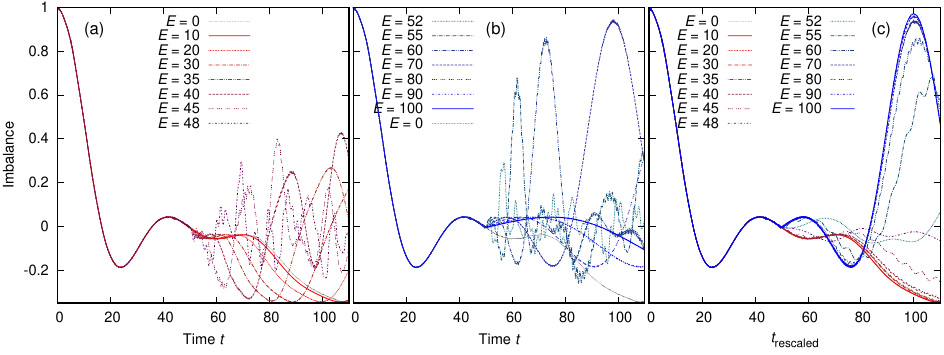}
    \caption{Time evolution of the local spin imbalance $\mathcal{I} = (-1)^{j-1}\mathcal{I}_{j}$ [See Eq.~\eqref{eq:imbalance} in the main text] for $L=10$ with the periodic boundary condition, before and after the field $E$ is switched on at $t_\mathrm{on} = 50$. We set $U=50$ and $\Delta t = 1/3200$, and choose the singly occupied state $\vert\uparrow\downarrow\uparrow\downarrow\cdots\uparrow\downarrow\rangle$ as the initial state. We vary $E$ for $0\leq E\leq 2U$ and the data for $E \leq U$ and $E \geq U$ are shown in the panel (a) and (b) respectively. In the panel (c), we show all the data with a rescaled time defined in Eq.~\eqref{eq:rescaled_time}. $t_h$ ($t_h^{-1}$) is used as the unit of energy (time).
    }
    \label{fig:imbalance_U50_PBC}
\end{figure*}

\section{Details of the numerical computation}
\label{sec:numerical}

\subsection{Fourth-order Runge-Kutta method}
\label{subsec:RK4}

We employ the fourth-order Runge-Kutta method for computing the real-time evolution of the wavefunction of the system according to the time-dependent Schr\"odinger equation
\begin{equation}
    i \frac{\partial}{\partial t}\vert\psi(t)\rangle = H(t)\vert\psi(t)\rangle.
    \label{eqn:Schroedinger}
\end{equation}
For time step $\Delta t$, the wavefunction at time $t + \Delta t$ is approximated by defining $\vert k_0\rangle = \mathbf{0}$ and computing
\begin{gather}
    \vert k_{j}\rangle = -i \Delta t H(t+\alpha_j\Delta t) \left(\vert \psi(t) \rangle - \beta_j i \Delta t \vert k_{j-1}\right)\rangle,
    \end{gather}
for $j=1,2,3,4$ and 
    \begin{gather}
    \vert \psi(t + \Delta t)\rangle = \vert \psi(t)\rangle
    + \sum_{j=1}^4 a_j \vert k_j\rangle,
    \end{gather}
with
$(\alpha_1, \alpha_2, \alpha_3, \alpha_4) = (0,1/2,1/2,1),
(\beta_1,\beta_2,\beta_3, \beta_4) = (0,1/2,1/2,1)$, and $
(a_1, a_2, a_3, a_4) = (1/6, 1/3, 1/3, 1/6)$.
We adopt the set of basis diagonal in $n_{j\sigma}$ for all $(j,\sigma)$.
For the half-filled system with zero net spin polarization, the Hilbert space dimension is $\begin{pmatrix}L\\L/2\end{pmatrix}^2$, while the matrix representation of the tight-binding Hamiltonian has at most $4L-3$ ($4L+1$) non-zero matrix elements for the open (periodic) boundary condition.

\subsection{Gauge choice}
\label{subsec:gaugechoice}
For the length gauge Hamiltonian \eqref{eq:model_lgauge}, the contribution from the site energy term to the diagonal matrix elements is a multiple of $E$, which can be $\mathcal{O}(10^3t_h)$ in our simulation.
This is significantly larger compared to the case of the velocity gauge \eqref{eq:model_vgauge}, with only the on-site interaction, which is at most $(L/2)U$, contributing to the diagonal matrix elements,
while the absolute value of the non-zero off-diagonal matrix elements is always $t_h$.

For the time evolution by the Runge-Kutta method to be accurate, the time step $\Delta t$ has to satisfy $\vert H\psi\vert \Delta t \ll 1$ for each normalized input state $\psi$, where $\vert\phi\vert \equiv \sqrt{\langle\phi\vert\phi\rangle}$.
We have confirmed that with sufficiently small values of $\Delta t$, the results for the two gauge choices agree completely for the open boundary condition, and do not significantly differ in the periodic boundary condition for $E, U\gg 1$. The latter is because when $E$ is large so that the hopping between any two neighboring sites is suppressed, the site energy difference of $(L-1)E$ in the length gauge between sites 1 and $L$ is even larger, therefore the hopping between these two sites is more strongly suppressed.

\begin{figure}
    \centering
    \includegraphics{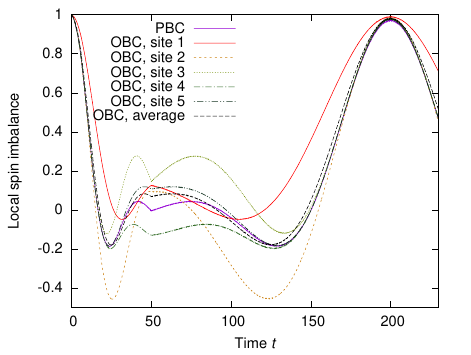}
    \caption{Local spin imbalance for $U=50$ before and after $E=100$ is switched on at $t_\mathrm{on}=50$. We set $\Delta t= 1/3200$ and use the velocity gauge. For the periodic boundary condition (PBC), $\mathcal{I}=(-1)^{j-1}\mathcal{I}_j$, which is $j$-independent with PBC, is plotted. For the open boundary condition (OBC), the local value $\tilde{\mathcal{I}}_{j}=(-1)^{j-1}\mathcal{I}_{j}$ ($j=1,2,\cdots, 5$) and the averaged value $\sum_{i=1}^5 \tilde{\mathcal{I}}_j /5$ are plotted. $t_h$ ($t_h^{-1}$) is used as the unit of energy (time).}
    \label{fig:imbalance_U50_PBC_OBC}
\end{figure}

\subsection{Boundary effect}
\label{subsec:boundary}
Here, we examine the effect of the boundary.
In Fig.~\ref{fig:imbalance_U50_PBC} we have plotted the local spin imbalance in the case of the \textit{periodic} boundary condition, while we have studied the open boundary condition in the main text. The settings studied in Fig.~\ref{fig:imbalance_U50_PBC} are the same as in Fig.~\ref{fig:imbalance} in the main text except for the boundary condition. While the details of the spin dynamics is different, the local spin imbalance overlaps with each other when plotted against the rescaled time both for $E\ll U$ (unless $U/E$ is an integer) and $E\gg U$.
In Fig.~\ref{fig:imbalance_U50_PBC_OBC} we have plotted the site-dependence of local spin imbalance. For the open boundary condition, the site dependence of the dynamics is significant, however they are reversed at the same time when the electric field is switched on at $t_\mathrm{on} = 50$.

\begin{figure}
\includegraphics[width=7.5cm]{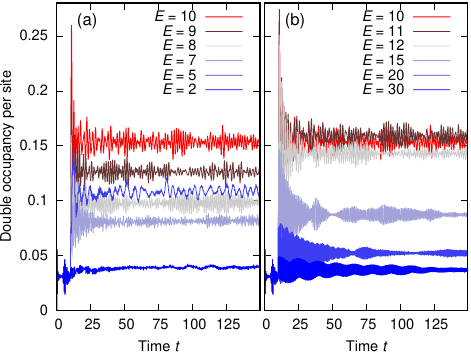}
\caption{Time evolution of the doublon number per site $\mathcal{N}_\mathrm{double}(t)$ [Eq.~\eqref{eq:double_occ} in the main text], where electric field $E$ is switched on at $t=50$. The data for $E\leq (\geq)~10$ are shown in the left (right) panel. We set $U=10$, $L=10$, and $\Delta t=1/1600$, and choose the singly occupied state $\vert\uparrow\downarrow\uparrow\downarrow\cdots\uparrow\downarrow\rangle$ as the initial state.
$t_h$ ($t_h^{-1}$) is used as the unit of energy (time). }
\label{fig:doubleoccU=10}
\end{figure}

\section{Numerical results for smaller values of $U$}
\label{sec:smallerU}

\begin{figure*}
    \centering
    \includegraphics{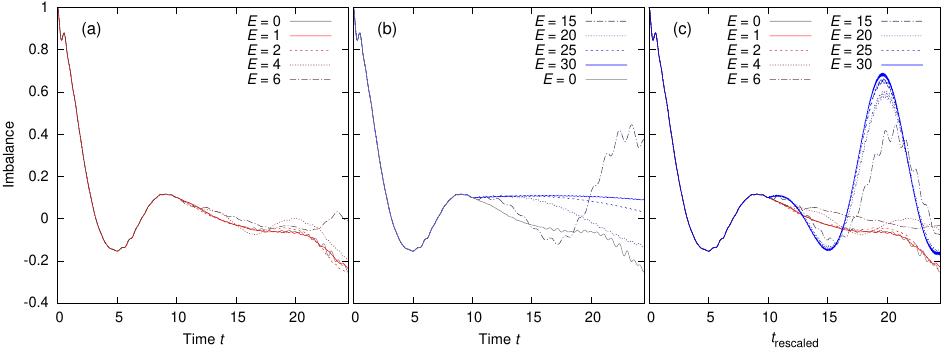}
    \caption{Time evolution of the local spin imbalance $\mathcal{I}_{L/2}$ [Eq.~\eqref{eq:imbalance} in the main text] for $L=10$ before and after the field $E$ is switched on at $t_\mathrm{on} = 10$. We set $U=10$ and $\Delta t=1/1600$, and choose the singly occupied state $\vert\uparrow\downarrow\uparrow\downarrow\cdots\uparrow\downarrow\rangle$ as the initial state. We vary $E$ for $0\leq E\leq 2U$ and the data for $E \leq U$ and $E \geq U$ are shown in the panel (a) and (b) respectively. In the panel (c), we show all the data with a rescaled time defined in Eq.~\eqref{eq:rescaled_time} in the main text. $t_h$ ($t_h^{-1}$) is used as the unit of energy (time).}
    \label{fig:imbalanceU=10}
\end{figure*}

Here we discuss the robustness of the effective spin picture presented in the main text down to smaller values of $U$, taking $U=10$ as an example.
In Fig.~\ref{fig:imbalanceU=10}, we plot the local spin imbalance at one of the center sites for $L=10$, with the singly occupied state as the initial state as in the main text (see Fig.~\ref{fig:imbalance}) and the electric field $E$ is switched on at $t=10$. Here we have excluded the cases with $E=5(=U/2)$ and $E\sim 10(=U)$, for which the doublon occupancy after $t=10$ becomes significant.
The change of the local spin imbalance is more rapid compared to the $U=50$ case, reflecting the larger value of $J_0(\propto 1/U)$. The plots against $t_\mathrm{rescaled}$ again overlap with each other, exhibiting dynamics similar to the $E=0$ case for $E<U$ and its reverse for $E>U$, though with larger discrepancies compared to the $U=50$ case.

In Fig.~\ref{fig:doubleoccU=10}, we plot the doublon number per site against time, for values of $E$ including $E=5$ and $E\sim 10$. While the quantity rapidly increases from zero to $\sim 0.03$ at the start of the dynamics reflecting the smaller cost of double occupancy compared to the $U=50$ case in the main text, the ranges of the values for $E=U/2$ and $E=U$ are comparable to those shown in Fig.~\ref{fig:doubleocc}. For $E\gtrsim15$, the double occupancy stays below around $0.1$.

\bibliographystyle{apsrev4-1}
\bibliography{ref.bib}

\begin{thebibliography}{71}
\bibitem [{FN1()}]{FN1}%
  \BibitemOpen
  \href@noop {} {}\bibinfo {note} {The numerical method is explained in Appendix~\ref{subsec:RK4}}\BibitemShut {NoStop}%
\end{thebibliography}
\end{document}